\documentclass[10pt,twocolumn,%
superscriptaddress,showkeys,showpacs,nofootinbib,byrevtex]{revtex4-1}
\usepackage{srcltx}
\usepackage{graphicx}
\usepackage{epsfig,bm,dcolumn}
\usepackage{epstopdf}
\usepackage{xcolor}
\usepackage{graphicx}
\usepackage{slashed}

\begin{document}

\title {Regge approach to the reaction of $\gamma N \to K^* \Lambda$}

\author{Byung-Geel Yu}
\email[ ]{bgyu@kau.ac.kr}
\affiliation{Research Institute of Basic Sciences, Korea
Aerospace University, Goyang, Gyeonggi 10540, Korea}

\author{Yongseok Oh}%
\email[]{yohphy@knu.ac.kr}
\affiliation{Department of Physics, Kyungpook National University,
Daegu 41566, Korea}
\affiliation{Asia Pacific Center for Theoretical Physics, Pohang,
Gyeongbuk 37673, Korea}

\author{Kook-Jin Kong}
\email[ ]{kong@kau.ac.kr}
\affiliation{Research Institute of Basic Sciences, Korea
Aerospace University, Goyang, Gyeonggi 10540, Korea}

\begin{abstract}
Photoproduction of $K^*$ vector mesons off nucleon is investigated
within the Regge framework where the electromagnetic vertex of
$\gamma K^*K^*$ fully takes into account the magnetic dipole and
electric quadrupole moments of spin-1 $K^*$ vector meson. The
$t$-channel $K^*(892)$, $K(494)$ and $\kappa(800)$ meson exchanges
are considered for the analysis of the production mechanism. The
experimentally observed rapid decrease of the cross sections for
the $\gamma p \to K^{*+} \Lambda$ reaction beyond the resonance
region is well reproduced by the dominance of the exchange of
$K$-meson trajectory. The role of the scalar $\kappa$-meson
trajectory is found to be minor in both $\gamma p$ and $\gamma n$
reactions. The cross sections for the $\gamma n \to K^{*0}
\Lambda$ reaction are predicted to be about twice those of the
$\gamma p \to K^{*+}\Lambda$ reaction. The role of the $K^*$
electromagnetic multipoles and the proton anomalous magnetic
moment is studied through the total and differential cross
sections and spin/parity asymmetries. We suggest the measurement
of the photon polarization asymmetry as a tool for identifying the
role of the magnetic dipole and electric quadrupole moments of the
$K^*$ vector meson.
\end{abstract}



\maketitle

Electromagnetic properties of hadrons are important to unravel the internal structure of hadrons.
In the case of vector mesons, in particular, their magnetic moments and electric quadrupole moments
are known to be $\mu = e_V^{}/m_V^{}$ and $\mathcal{Q} = - e_V^{}/m_V^2$~\cite{KT73,BH92},
respectively, in the limit of point-like structure, where $e_V^{}$ ($m_V^{}$) is the charge
(mass) of the vector meson.
Therefore, any deviation from these canonical values implies nontrivial internal structure of
vector meson and may give a clue on the properties of the constituents of vector mesons~\cite{DDM62}.

In this respect, the recent experimental data from the CLAS
Collaboration on $K^*$ photoproduction at the Thomas Jefferson
National Accelerator Facility~\cite{CLAS13,Mattione14} draw
attention as they provide information on the properties of vector
mesons as well as on the production mechanisms of strangeness via
the spin-1 vector meson. Nevertheless, only a few model
calculations were attempted to analyze the reaction processes
$\gamma p \to K^{*+} \Lambda$~\cite{OK06a,KNOK11b,ONH09} and
$\gamma n \to K^{*0} \Lambda$~\cite{JUNHE15}. Moreover, the former
model calculations considered the electromagnetic interaction of
the $K^*$ vector meson with the charge coupling only by dropping
out other couplings. As stated above, however, spin-1 vector
mesons have non-vanishing magnetic dipole and electric quadrupole
moments. Therefore, investigating static properties of vector
mesons through their production mechanisms is desirable and it is
the main motivation of the present work.

In the recent publication~\cite{YK16}, two of us studied
photoproduction of charged $\rho$ meson, i.e., $\gamma N \to
\rho^\pm N$ including the electromagnetic multipoles of vector
mesons. We found that the existing data of
Refs.~\cite{BDLM79,AABH76,ABHHM74} on these processes could be
reasonably reproduced. Encouraged by this observation, in the
present work, we study the reaction processes $\gamma p \to K^{*+}
\Lambda$ and $\gamma n \to K^{*0} \Lambda$ in order to see the
role of the structure of the $\gamma K^*K^*$ vertex for a
description of existing data reported in
Refs.~\cite{CLAS13,Mattione14,ABHHM76}.

As discussed in Ref.~\cite{YK16},  the validity of the Ward
identity for the $\gamma K^*K^*$ vertex is crucial to provide a
reliable prescription for gauge invariance in charged-meson
photoproduction. The most general form of the electromagnetic
$\gamma K^*K^*$  vertex $\Gamma^{\mu\nu\alpha}_{\gamma
K^*K^*}(q,Q)$ which satisfies the Ward identity is given by
\begin{widetext}
\begin{eqnarray}\label{gvv}
\eta_\nu^* \Gamma^{\mu\nu\alpha}_{\gamma K^*K^*}(q,Q)
\eta_\alpha\epsilon_\mu
&=& - \eta_\nu^*(q) \Biggl\{ e_{K^*}^{} \left[(q+Q)^\mu g^{\nu\alpha}
- Q^\nu g^{\mu\alpha}-q^\alpha g^{\mu\nu}\right]
+ e \kappa_{K^*}^{} (k^\nu g^{\mu\alpha}- k^\alpha g^{\mu\nu})
\nonumber \\ && \mbox{}
- e \frac{(\lambda_{K^*}+\kappa_{K^*})}{2m^2_{K^*}}
\left[(q+Q)^\mu k^\nu k^\alpha - \frac12 \left( q+Q \right) \cdot k
\left( k^\nu g^{\mu\alpha}+k^\alpha g^{\mu\nu} \right) \right] \Biggr\}
\, \eta_\alpha(Q)\epsilon_\mu\,,
\end{eqnarray}
\end{widetext}
where $k_\mu$ is the photon momentum and $Q_\mu$ and $q_\mu$ are
the incoming and outgoing $K^*$ momenta, respectively, with $Q_\mu
= q_\mu - k_\mu$. The polarization vectors of the photon and $K^*$
are denoted by $\epsilon_\mu$ and $\eta_\mu$, respectively. Here
$e_{K^*}^{}$ is the charge of the $K^{*}$ vector meson, and the
magnetic dipole and electric quadrupole moments of the $K^*$ are
given as
\begin{eqnarray}
\mu_{K^*}^{} &=& \left(\tilde{e}_{K^*}^{}+\kappa_{K^*}^{} \right) \left( \frac{e}{2m_{K^*}} \right) ,
\nonumber \\
\mathcal{Q}_{K^*} &=& \lambda_{K^*} \left( \frac{e}{m^2_{K^*}} \right).
\end{eqnarray}
Theoretical estimates on the magnetic dipole and electric quadrupole moments of the $K^*$
vector meson are reported in various models inspired by QCD~\cite{HP98a,BM06}.
In the present work, we adopt the values predicted in Ref.~\cite{BM06}, namely,
$\tilde{e}_{K^*}^{}=+1$, $\kappa_{K^*}^{}=1.23$ (therefore, $\mu_{K^*}^{} = 2.23$)
and $\lambda_{K^*}^{}=-0.38$ for the $K^{*+}$ and
$\tilde{e}_{K^*}=0$, $\kappa_{K^*}=-0.26$ (therefore, $\mu_{K^*}^{} = -0.26$) and
$\lambda_{K^*}=0.01$ for the $K^{*0}$.
We also use $g^V_{K^*N\Lambda} = -4.5$ and $g_{K^*N\Lambda}^T = -10$ obtained
by using the flavor SU(3) relations with the ratios $\alpha_V^{}=1$ and $\alpha_T^{}=0.4$ from
$g^V_{\rho NN} = 2.6$ and $g^T_{\rho NN} = 9.62$ following Refs.~\cite{YK16,WF88}.

We now consider the $t$-channel Regge-pole exchange in production amplitudes.
Given the Born amplitude for the process $\gamma(k) + N(p) \to K^*(q) + Y(p')$, where the
momentum of each particle is denoted in the parenthesis, the meson exchange in the $t$-channel
is reggeized by replacing the $t$-channel pole with the Regge-pole in the form of
\begin{equation}
\mathcal{R}^{\varphi}(s,t) =
\frac{\pi\alpha'_J}{\Gamma[\alpha_J^{} (t)+1-J]\sin[\pi\alpha_J^{} (t)]}
\left( \frac{s}{s_0^{}} \right)^{\alpha_J^{}(t)-J}
\end{equation}
written collectively for a meson $\varphi$ of spin-$J$ with the phase
$\frac12 [(-1)^J+e^{-i\pi\alpha_J^{}(t)}]$ assigned to the exchange-nondegenerate single meson.

Recalling that the energy dependence of total cross sections is
given as $\sigma \sim s^{\alpha_J^{}(0)-1}$, the steep decease of
the cross section for the $\gamma p \to K^{*+} \Lambda$ reaction
with increasing photon energy, as observed by the CLAS
Collaboration~\cite{CLAS13}, implies the dominance of the exchange
of the kaon trajectory, whereas $K^*$ of nonzero spin and $K_2^*$
as well should be suppressed in the region over the resonance
peak.

\begin{widetext}
With these in mind we write the production amplitude as consisting of kaon, and scalar meson $\kappa$
in addition to the elementary
Born terms for gauge-invariant $K^*$ exchange, i.e.,
\begin{equation}
\mathcal{M} = \bar{u}(p') \, \eta^*_{\nu}(q) \left( \mathcal{M}_{K^*N}^{\mu\nu}
 + \mathcal{M}_K^{\mu\nu} + \mathcal{M}_\kappa^{\mu\nu} \right) \epsilon_\mu(k) \, u(p) \, ,
\end{equation}
where
\begin{eqnarray}
\mathcal{M}_{K^*N}^{\mu\nu} &=&
\Biggl\{ \left(g^V_{K^*N\Lambda} \gamma^\nu
+ \frac{g^T_{K^*N\Lambda}}{4M_N^{}}  \left[ \gamma^\nu, \slashed{q} \right] \right)
\frac{\slashed{p} +\slashed{k} + M_N}{s - M_N^2}
\left(e_N^{} \gamma^\mu - \frac{e\kappa_N^{}}{4M_N^{}}
\left[ \gamma^\mu, \slashed{k} \right] \right)
\nonumber\\
&& \quad \mbox{}
+ \left( e_Y^{} \gamma^\mu
- \frac{e\kappa_Y^{}}{4M_N^{}} \left[ \gamma^\mu, \slashed{k} \right] \right)
\frac{\slashed{p}' - \slashed{k} + M_Y^{} }{u - M_Y^2}
\left( g^V_{K^*N\Lambda} \gamma^\nu
+ \frac{g^T_{K^*N\Lambda}}{4M_N^{}} \left[ \gamma^\nu, \slashed{q} \right]
\right) \nonumber \\
&& \quad \mbox{}
+ \Gamma_{\gamma K^* K^*}^{\mu\nu\alpha}(q,Q)
\frac{-g^{\alpha\beta} + Q^\alpha Q^\beta / m^2_{K^*}}{t-m^2_{K^*}}
\left(g^V_{K^*N\Lambda}\gamma^\beta + \frac{g^T_{K^*N\Lambda}}{4M_N^{}}
\left[  \gamma^\beta,\slashed{Q} \right]\right)
\nonumber\\ && \quad \mbox{}
- \left( e_N^{} - e_{Y}^{} \right)
\frac{g_{K^*N\Lambda}^T}{4M_N^{}} \left[ \gamma^\nu, \gamma^\mu \right]
\Biggr\}
\left( t - m^2_{K^*} \right)
\mathcal{R}^{K^*}(s,t) \frac12 \left(-1+e^{-i\pi\alpha^{}_{K^*}(t)} \right),
\label{vector} \\
\mathcal{M}_K^{\mu\nu} &=&
i \, \frac{g_{\gamma KK^*}^{}}{m_0^{}} \, g_{KN\Lambda}^{} \,
\varepsilon^{\mu\nu\alpha\beta} k_\alpha^{} Q_\beta^{} \gamma_5^{} \mathcal{R}^K(s,t)
\left\{ \begin{array}{c} e^{-i\pi\alpha^{}_K(t)} \\ 1 \end{array} \right\},
\label{kaon} \\
\mathcal{M}_\kappa^{\mu\nu} &=& \frac{g_{\gamma \kappa K^*}^{}}{m_0^{}} \,
g_{\kappa N\Lambda}^{} \left( k \cdot Q \, g^{\mu\nu} -  Q^\mu k^\nu \right)
\mathcal{R}^\kappa(s,t)
\frac12 \left( 1+e^{-i\pi\alpha^{}_\kappa(t)} \right)
\label{scalar}
\end{eqnarray}
\end{widetext}
with the given phases of the Regge poles. We use the trajectories
for $K$, $\kappa$ and $K^*$ ~\cite{YCK11}
\begin{eqnarray}\label{pitraj}
\alpha_K^{}(t) &=& 0.7 \left( t-m^2_K \right)\,,\nonumber\\
\alpha_{\kappa}^{}(t) &=& 0.7 \left( t-m^2_\kappa \right) ,\nonumber\\
\alpha_{K^*}^{} (t) &=& 0.83\,t +0.25.
\end{eqnarray}
The nucleon and hyperon masses are denoted by $M_N$ and $M_Y$,
respectively, and $m_0^{} = 1~\mbox{GeV}$ is the mass scale
parameter.

For the $\gamma p\to K^{*+}\Lambda$ process, the proton pole in the $s$-channel and
the contact term are included for gauge-invariance of the $t$-channel $K^{*+}$ exchange.
For the case of the $\gamma n \to K^{*0}\Lambda$ process, however, only the $t$-channel
$\kappa+K+K^{*0}$ exchanges are included with the magnetic dipole and electric quadrupole
moments of $K^{*0}$ which are by themselves gauge-invariant.
In Eq.~(\ref{kaon}), the phase of the $K$ exchange is to be read for the
$\gamma p \to K^{*+} \Lambda$ (upper) and $\gamma n \to K^{*0}\Lambda$ (lower) in consistent
with the phase relations in the $\rho^\pm$ and $\pi^\pm$ photoproductions~\cite{YK16,YCK11a}.
In the reggeization of $K^{*+}$ exchange in the $t$-channel we preserve the proton magnetic
moment term in the $s$-channel Born term for the $\gamma p$ reaction because of the expected role of
the magnetic interactions between the particles of non-zero spin.

For the estimate of the kaon-trajectory exchange, we use $g_{KN\Lambda}^{} =-13.24$
in consistent with the SU(3) prediction with $\alpha=0.365$ and $g_{\pi NN}^{} = 13.4$,
and the couplings $g_{\gamma KK^{*+}}^{} =0.254$ and $g_{\gamma KK^{*0}}^{}=-0.388$ are
determined from the measured decay widths $\Gamma_{K^*\to K^+\gamma}=50.3$~keV and
$\Gamma_{K^*\to K^0\gamma}=116.6$ keV, respectively.
The negative sign for $g_{\gamma KK^{*0}}^{}$ follows the quark model prescription.

The radiative decay constants relevant to the scalar meson $\kappa$ is unknown at present
and we use the prediction of Ref.~\cite{BHS02}, which
gives the decay width of $K^{*0}$ as
\begin{eqnarray}
\Gamma(K^{*0}\to \kappa\gamma) = \frac{1}{96\pi} {e^2\over
\tilde{g}_\rho^2} \left({m_{K^*}^2-m_\kappa^2\over m_{K^*}}
\right)^3 \left|-{8\over3} \beta_A \right|^2,
\end{eqnarray}
supposing that the $K^*$ mass larger than the $\kappa$ mass. The
values of $\beta_A=0.72~\mbox{GeV}^{-1}$ and $\tilde{g}=4.04$ are
estimated in Ref.~\cite{BHS02}. In this work we consider
$m_\kappa=800$~MeV and $g_{\gamma \kappa K^{*0}} =
0.144~\mbox{GeV}^{-1}$, which gives $\Gamma(K^{*0}\to \kappa
\gamma) \approx 0.411$~keV. The SU(3) relation $g_{\gamma\kappa
K^{*0}} = -2g_{\gamma\kappa K^{*\pm}}$ is kept through the present
work, which gives $g_{\gamma \kappa K^{*+}}=-0.072$~\cite{OK06b}.
For the scalar meson-baron coupling constant we adopt $g_{\kappa
N\Lambda}=-14.7$ using a recent result of QCD sum rule
calculation~\cite{ETOR06}.

\begin{figure}[b]
\includegraphics[width=0.99\columnwidth]{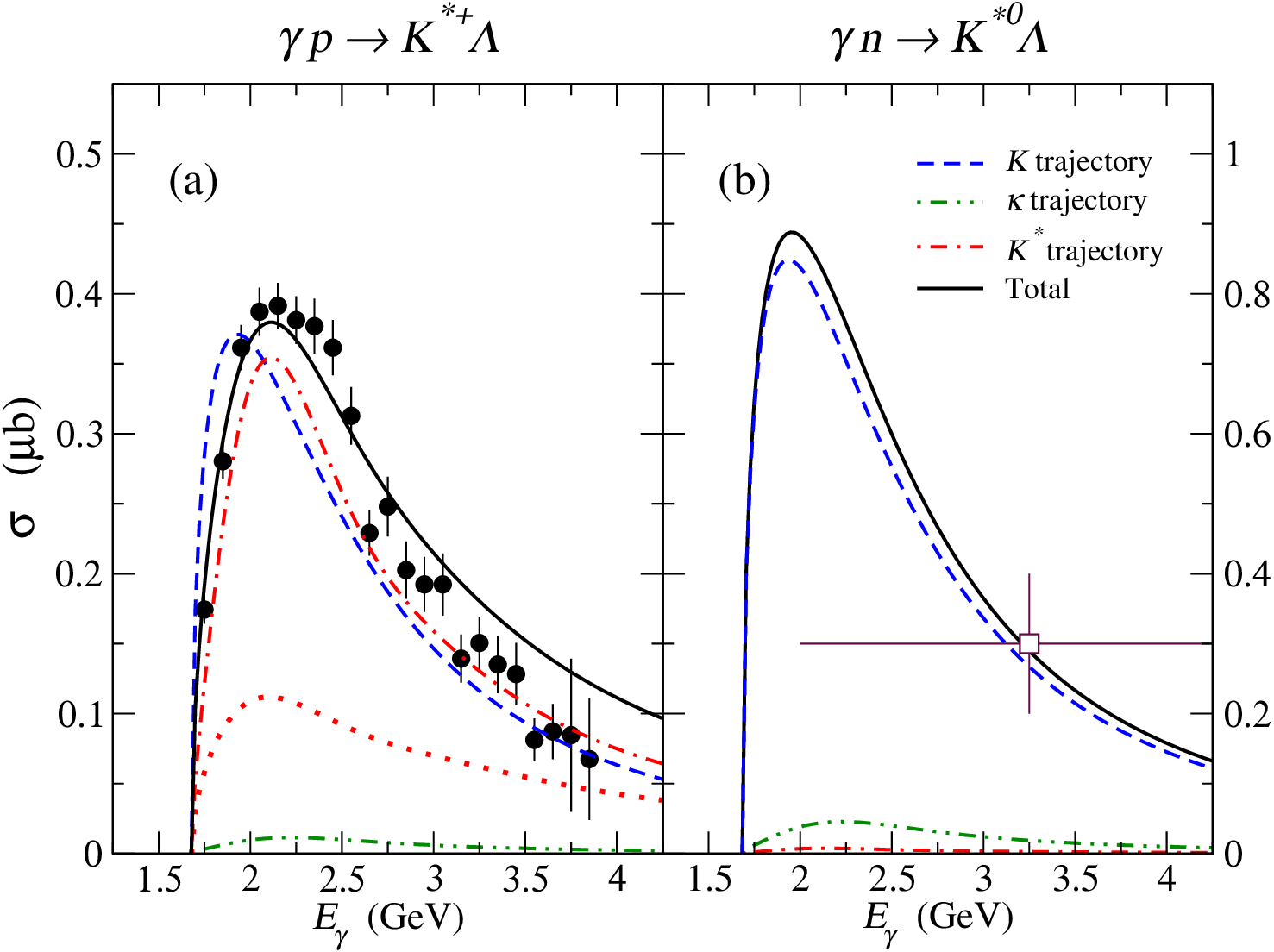}
\caption[]{ Total cross section (a) for $\gamma p \to K^{*+}
\Lambda$ and (b) for $\gamma n \to K^{*0} \Lambda$. Blue dashed,
green dot-dot-dashed, and red dot-dashed  lines are the
contributions from the exchanges of $K$, $\kappa$, and $K^*$ with
$(\kappa_{K^*}^{},\,\lambda_{K^*}^{})=(1.23,-0.38)$, respectively.
The gauge-invariant $K^*$ exchange $\mathcal{M}_{K^*N}$ in
Eq.~(\ref{vector}) is given by the red dotted line. The solid
lines show the results of the full calculation. Experimental data
for the $\gamma p$ reaction are from Ref.~\cite{CLAS13} (filled
circles) and those for the $\gamma n$ reaction are from
Ref.~\cite{ABHHM76} (open square).} \label{fig1}
\end{figure}

Shown in Fig.~\ref{fig1} are the total cross sections for the
reactions of $\gamma p \to K^{*+} \Lambda$ and $\gamma n \to
K^{*0} \Lambda$ as functions of the photon energy $E_\gamma$ in
the laboratory frame. The contributions from $K$, $\kappa$ and
$K^*$ exchanges are displayed by the dashed, dot-dot-dashed and
dot-dashed lines in order. The red dotted line is from the
gauge-invariant $K^*$ exchange $\mathcal{M}_{K^*N}$ in
Eq.~(\ref{vector}) which contains the $s$-channel proton pole and
contact terms. The experimental data are from the CLAS
collaboration~\cite{CLAS13} for $\gamma p \to K^{*+} \Lambda$ and
from the ABHHM collaboration~\cite{ABHHM76} for $\gamma n \to
K^{*0} \Lambda$, respectively. These results show that the
$K^*\Lambda$ production in the $\gamma p$ reaction exhibits the
dominance of $K$ plus $K^*$ exchanges which are comparable to each
other, while the $\gamma n$ process is totally governed by the $K$
exchange. As a result, the cross section for the $\gamma n$
reaction is about double the size of the $\gamma p$ cross section.
But the effect of $K^*$ electromagnetic multipoles are suppressed
in the $\gamma n$ reaction.

\begin{figure}[t]
\includegraphics[width=0.95\columnwidth]{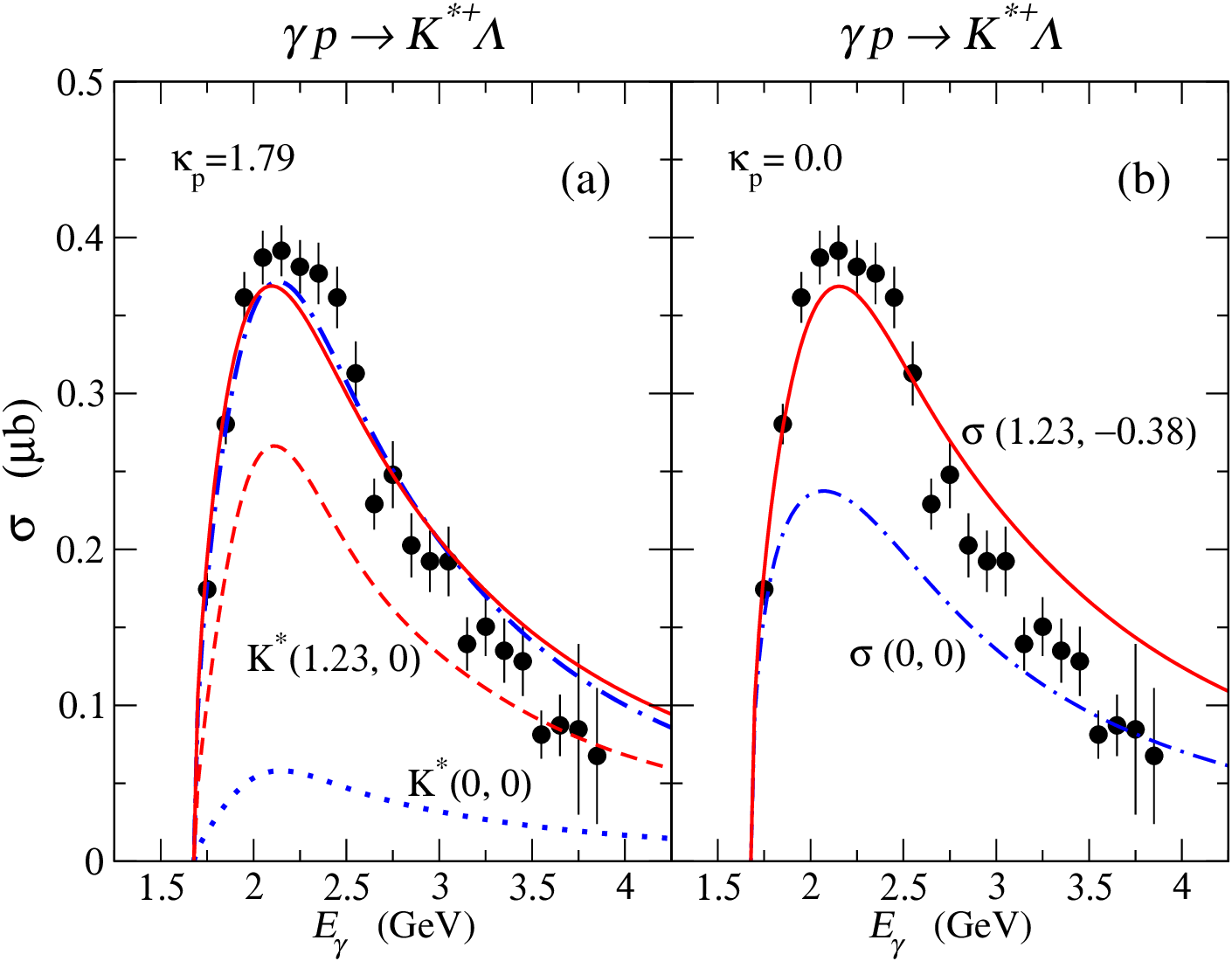}
\caption[]{ Total cross sections for $\gamma p \to K^{*+} \Lambda$
(a) with and (b) without $\kappa_p$. In (a), the red dashed line
is the $K^*$ exchange contribution with
$(\kappa_{K^*}^{},\,\lambda_{K^*}^{})=(1.23,0)$ and the
corresponding cross section is given by the red solid line. The
blue dash-dotted line is the cross section from the $K^*$ exchange
with $(\kappa_{K^*}^{},\,\lambda_{K^*}^{})=(0, 0)$ which is given
by blue dotted line. In (b), full calculations for cross sections
with $(\kappa_{K^*}^{},\,\lambda_{K^*}^{})=(1.23,-0.38)$ and
$(0,0)$ are shown by the same notation as in (a), respectively. }
\label{fig2}
\end{figure}

As stated before, since the process involves the production of
spin-1 vector meson, the effects of magnetic interactions are
expected. In the present work, we examine the role of the
electromagnetic multipoles of $K^*$ as well as of the proton pole
term in the case of the $\gamma p$ process. Figure~\ref{fig2}
shows the role of the $\kappa_{K^*}^{}$ and $\lambda_{K^*}^{}$
terms in the cross sections for the $\gamma p$ process with and
without the proton anomalous magnetic moment $\kappa_p^{}$ term.
We adopt $(\kappa_{K^*}^{},\,\lambda_{K^*}^{})=(1.23,-0.38)$
following Ref.~\cite{BM06}. We first note that the difference of
the cross section $\sigma$ between the cases of
$(\kappa_{K^*}^{},\,\lambda_{K^*}^{})=(1.23,-0.38)$ by the solid
line and $(\kappa_{K^*}^{},\,\lambda_{K^*}^{})=(0,0)$ by the
dash-dotted line is noticeable in the case of $\kappa_p^{} = 0$ as
shown in Fig.~\ref{fig2}(b). This tendency disappears, however, in
the presence of $\kappa_p^{} = 1.79$ as can be seen in
Fig.~\ref{fig2}(a), which shows a nontrivial role of the proton
anomalous magnetic moment $\kappa_p$. Therefore, this signifies
that the $\kappa_p$ in the proton pole term should be activated in
the reggeization of the production amplitude in case of the vector
meson not only for a theoretical consistency but also for the
phenomenological consequences just we have demonstrated.

\begin{figure}[t]
\includegraphics[width=\columnwidth]{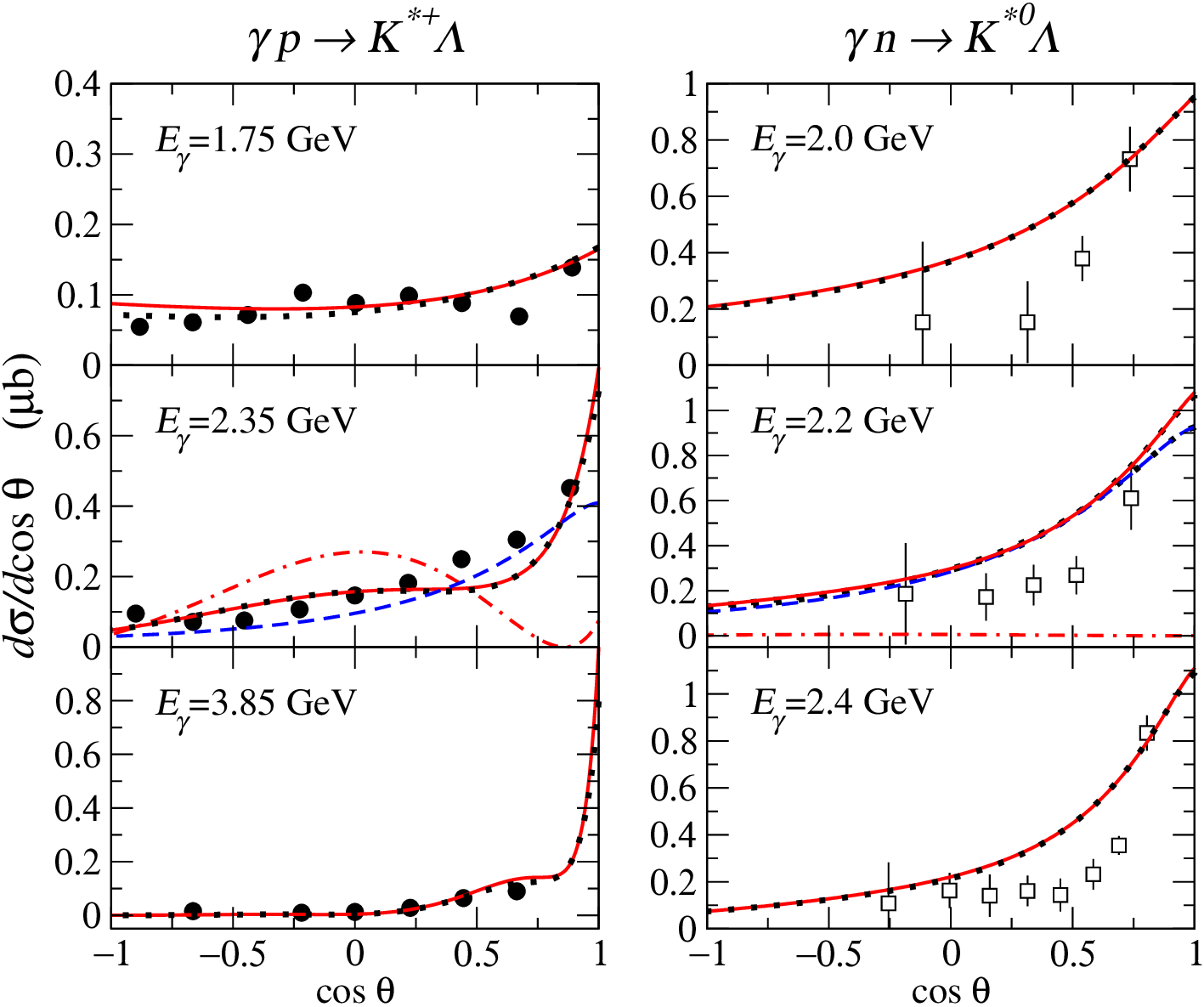}
\caption[]{ Differential cross sections for $\gamma p \to K^{*+} \Lambda$ and
$\gamma n \to K^{*0} \Lambda$ reactions.
Dotted lines are the results without $\kappa_{K^*}^{}$ and $\lambda_{K^*}^{}$ terms.
The respective contributions of $K$ and $K^*$ exchanges are shown with the same
notation as in Fig.~\ref{fig1}.
Experimental data are taken from Ref.~\cite{CLAS13} (filled circles) and
from Ref.~\cite{Mattione14} (open squares). }
\label{fig3}
\end{figure}

The observed differential cross sections for both reaction
processes are reasonably reproduced by the present model
calculations as shown in Fig.~\ref{fig3}. Enhancement at forward
angles in both cases illustrates the dominance of $K$ exchange
shown by the blue dashed lines in Fig.~\ref{fig3} for $E_\gamma =
2.35$~GeV (2.2~GeV) in the case of $\gamma p$ ($\gamma n$)
reaction. As in the case of total cross section, the contribution
of $K^*$ without multipoles does not significantly alter the
differential cross section in the presence of $\kappa_p$.

\begin{figure}[t]
\includegraphics[width=\columnwidth]{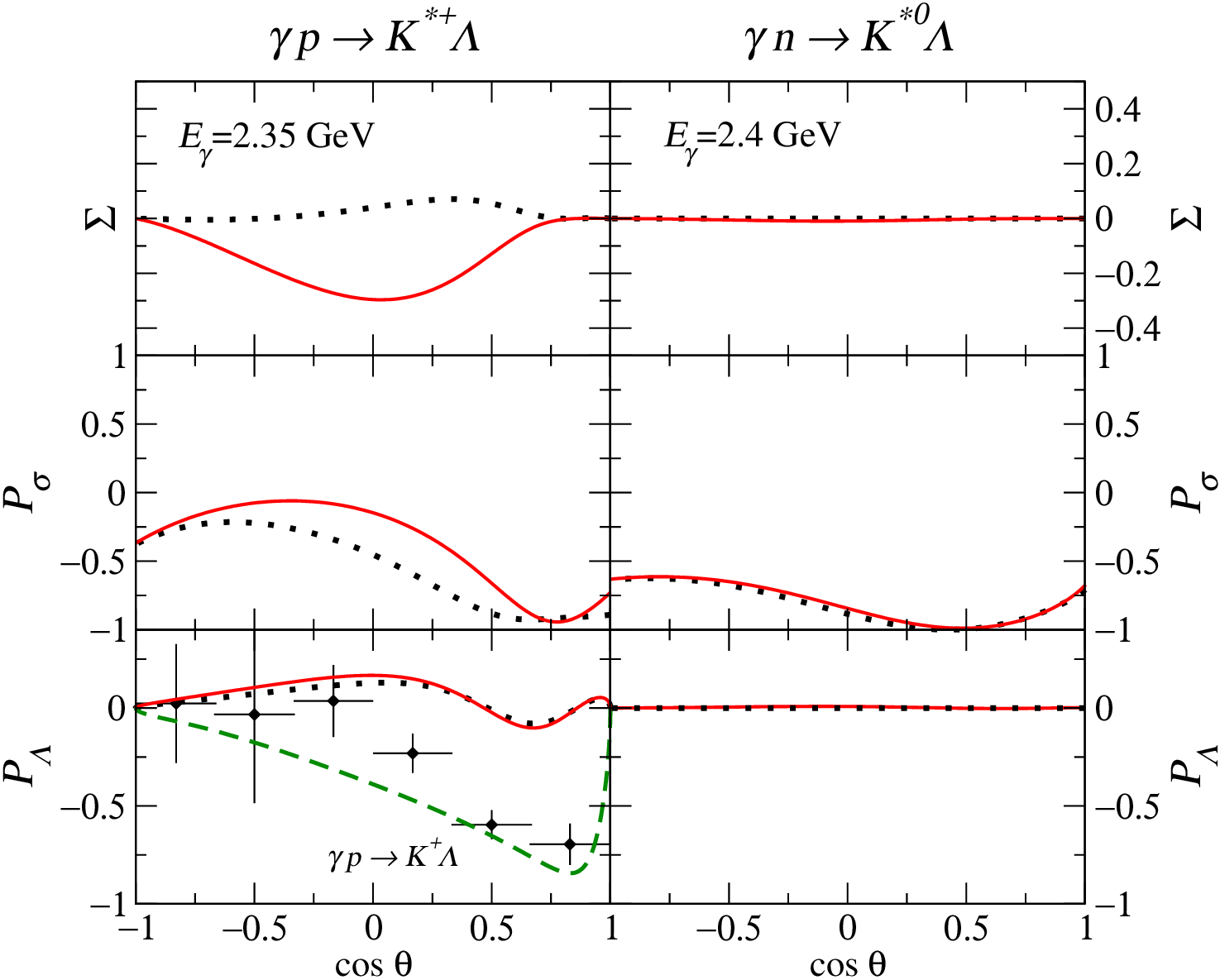}
\caption[]{
Spin observables of $\gamma p \to K^{*+}\Lambda$ at $E_\gamma=2.35$~GeV and
$\gamma n \to K^{*0}\Lambda$ at $E_\gamma=2.4$~GeV.
The solid lines are from the full calculation while the dotted lines are obtained without
$\kappa_{K^*}^{}$ and $\lambda_{K^*}^{}$ terms.
For comparison, the recoil polarization for $\gamma p\to K^+\Lambda$ is given together
by the green dashed line (left) with data taken from Ref.~\cite{GBBH03}.}
\label{fig4}
\end{figure}

Although contributions of $K^*$ multipoles, i.e.,
$\kappa_{K^*}^{}$ and $\lambda_{K^*}^{}$ terms, are small in cross
sections, their effects may be found in spin asymmetries.
Furthermore, since the contributions of $N^*$ resonances in the
$\gamma p \to K^{*+} \Lambda$ process are found to be rather
insignificant~\cite{KNOK11b}, the process is of benefit to the
measurement of such observables on a clean background. Presented
in Fig.~\ref{fig4} are the photon polarization observable
($\Sigma$) together with the parity asymmetry ($P_\sigma$) and the
recoil polarization ($P_\Lambda$) for both processes. Following
the convention and definitions of Refs.~\cite{PST96,TOYM98}, the
photon polarization asymmetry is given by
\begin{eqnarray}
\Sigma={\sigma^{(\perp,0,0,0)}-\sigma^{(\parallel,0,0,0)}\over
\sigma^{(\perp,0,0,0)}+\sigma^{(\parallel,0,0,0)}} ,
\label{ba}
\end{eqnarray}
where we define $\sigma^{(B,T,Y,V)}$ for the differential cross section $d\sigma /d\Omega$, where
the superscripts $(B,T,Y,V)$ denote the polarizations of  photon beam, target proton, produced
hyperon, and produced $K^*$ vector-meson, respectively.
The superscript $0$ means unpolarized state and $\parallel$ ($\perp$) corresponds to a photon
linearly polarized parallel (perpendicular) to the reaction plane but normal to the photon beam
direction.
The negativeness of $\Sigma$ in $\gamma p$ reaction is largely due to the $K^*$-exchange, which
reveals a sizable dependence on $\kappa_{K^*}^{}$ and $\lambda_{K^*}^{}$.
Thus, measuring this observable is desirable to verify the role of the $K^*$ magnetic moment and
electric quadrupole moment.

The parity asymmetry defined as
\begin{eqnarray}
P_\sigma = 2\rho^1_{1-1} - \rho^1_{00}\,,
\end{eqnarray}
measures the asymmetry between the natural and unnatural parity of
exchanged mesons in terms of density matrix elements~\cite{SSW70}.
In particular, the predicted value $P_\sigma \approx -1$ for the
$\gamma n$ reaction is understood by the dominance of the $K$
exchange of the unnatural-parity over the natural-parity exchanges
of $\kappa$ and $K^*$. Recently $P_\sigma$ of the $\gamma p \to
K^{*0} \Sigma^+$ reaction was reported~\cite{LEPS11} and the
comparison of $P_\sigma$ in various channels for the $K^*$
production will be useful to understand the production mechanism
of strangeness via the spin-1 vector meson.

The observation of the recoil polarization, $P_\Lambda$, is also
interesting since the produced $\Lambda$ hyperon in the final
state is self-analyzing~\cite{GBBH03}. The asymmetry between the
spin polarizations of the final $\Lambda$ along with the $y'$-axis
as defined by~\cite{PST96}
\begin{eqnarray}
P_\Lambda = \frac{\sigma^{(0,0,+y’,0)}-\sigma^{(0,0,-y’,0)}}
{\sigma^{(0,0,+y’,0)}+\sigma^{(0,0,-y’,0)}} ,
\label{recoil}
\end{eqnarray}
which can be measured by the subsequent weak decay of the final
$\Lambda$ through the $\Lambda \to p \pi$ decay. Viewed from the
different spin structure of the final state $K^{*+} \Lambda$ from
the case of the $K^+\Lambda$ in the final state of the reaction
process $\gamma p \to K^+ \Lambda$, it is informative to compare
$P_\Lambda$ asymmetry in both reactions. The dashed line for the
$P_\Lambda$ in the latter process is estimated from
Ref.~\cite{YCK11} with the experimental data of
Ref.~\cite{GBBH03}, which shows a quite different spin
polarization of $\Lambda$ from the $K^{*+} \Lambda$ production
process, as expected. The experimental measurements of these
observables, therefore, will be a testing ground to confirm models
for the electromagnetic production of strangeness through spin-1
vector mesons with electromagnetic multipoles.

In summary, we have investigated photoproduction reactions of $\gamma N \to K^*\Lambda$
focusing on the role of electromagnetic multipoles of $K^*$ vector meson.
With the $\gamma K^*K^*$ vertex that fully accounts magnetic dipole and electric quadrupole
moments of $K^*$ satisfying the Ward identity, analysis of existing data was performed based
on the Regge approach.
We found that the $\gamma p \to K^{*+} \Lambda$ process is dominated by $K$ plus $K^*$
exchanges, while $\gamma n \to K^{*0}\Lambda$ process can be understood by the
dominance of $K$ exchange.
Although the dependence of total and differential cross sections on the $K^*$ electromagnetic
multipoles cannot be easily verified, we found that the photon beam asymmetry may be useful
to unravel the role of $K^*$ multipoles in the production of $K^*$ vector meson.
Our predictions for spin observables would be tested by future measurements at current
electron/photon beam facilities.

\section*{Acknowledgments}

This work was supported by Basic Science Research Program through
the National Research Foundation of Korea (NRF) funded by the
Ministry of Education (Grant Nos. NRF-2013R1A1A2010504 and
NRF-2015R1D1A1A01059603).

\end{document}